\documentclass[epj]{svjour} 
\usepackage[numbers,sort&compress]{natbib}
\usepackage{times}

\newcommand{\affiliation}[1]{\inst{#1}}
\newcommand{\ISKP}{1}
\newcommand{\HH}{2}
\newcommand{\IKP}{3}
\newcommand{\AISKP}{Helmholtz-Institut f\"ur Strahlen- und Kernphysik,
Universit\"at Bonn, D-53115 Bonn, Germany}
\newcommand{\AHH}{Institut f\"ur Experimentalphysik, Universit\"at Hamburg,
D-22761 Hamburg, Germany}
\newcommand{\AIKP}{Institut f\"ur Kernphysik, Forschungszentrum J\"ulich,
D-52425 J\"ulich, Germany}

\newcommand{\authors}{
        \author{
{M.~Altmeier} \affiliation{\ISKP} 
\and{F.~Bauer} \affiliation{\HH}
\and{J.~Bisplinghoff} \affiliation{\ISKP} 
\and{K.~B\"u\ss{}er} \affiliation{\HH} 
\and{M.~Busch} \affiliation{\ISKP} 
\and{T.~Colberg} \affiliation{\HH}  
\and{L.~Demir\"ors} \affiliation{\HH} 
\and{H.P.~Engelhardt} \affiliation{\ISKP}
\and{P.D.~Eversheim} \affiliation{\ISKP} 
\and{K.O.~Eyser} \affiliation{\HH} 
\and{O.~Felden} \affiliation{\IKP}  
\and{R.~Gebel} \affiliation{\IKP} 
\and{M.~Glende} \affiliation{\ISKP} 
\and{J.~Greiff} \affiliation{\HH} 
\and{F.~Hinterberger} \affiliation{\ISKP}
\and{E.~Jonas} \affiliation{\HH} 
\and{H.~Krause} \affiliation{\HH}
\and{T.~Lindemann} \affiliation{\HH}
\and{J.~Lindlein} \affiliation{\HH}
\and{B.~Lorentz} \affiliation{\IKP} 
\and{R.~Maier} \affiliation{\IKP} 
\and{R.~Maschuw} \affiliation{\ISKP}
\and{A.~Meinerzhagen} \affiliation{\ISKP}
\and{D.~Prasuhn} \affiliation{\IKP}
\and{H.~Rohdje\ss}\affiliation{\ISKP}
\and{D.~Rosendaal} \affiliation{\ISKP}
\and{P.~von~Rossen} \affiliation{\IKP}
\and{N.~Schirm} \affiliation{\HH}
\and{V.~Schwarz} \affiliation{\ISKP}
\and{W.~Scobel} \affiliation{\HH} 
\and{H.-J.~Trelle} \affiliation{\ISKP}
\and{K.~Ulbrich} \affiliation{\ISKP}
\and{E.~Weise} \affiliation{\ISKP}
\and{A.~Wellinghausen} \affiliation{\HH}
\and{R.~Ziegler} \affiliation{\ISKP}
}
} 

\newcommand{\Beginabstract}{\abstract}
\newcommand{\Endabstract}{}
\newcommand{\collaboration}[1]{}
\newcommand{\mypacs}{
\PACS{
        {25.40.Cm}{Elastic proton scattering} \and
        {13.75.Cs}{Nucleon-nucleon interactions} \and
        {24.70.+s}{Polarization phenomena in reactions} \and
        {21.30.-x}{Nuclear forces}
}
}

\newcommand{\institutes}{
\institute{\AISKP} \and 
\institute{\AHH} \and 
\institute{\AIKP} }
\renewcommand{\institutes}{
\institute{\AISKP \and \AHH \and \AIKP}
\offprints{F.~Hinterberger, HISKP, Universit\"at Bonn, Nussallee 14-16,
        D-53115 Bonn, Germany}
\mail{hinterberger@hiskp.uni-bonn.de}}
\newcommand{\preprint}[1]{\titlerunning{#1}}
\newcommand{\printfigures}{}

\newcommand{\Beginack}{\begin{acknowledgement}}
\newcommand{\Endack}{\end{acknowledgement}}
\newcommand{\figwidth}{8.5cm}

\renewcommand{\figwidth}{\columnwidth}

\newcommand{\begindescription}[1]{\begin{description}[#1]}

\usepackage{graphicx}
\usepackage{amssymb}

\newcommand{\MacroRev}{$Revision: 1.1 $}

\newcommand{\mylabel}[1]{\label{#1}
       \makebox[0cm][r]{\raisebox{5mm}[0mm][0mm]{\framebox{\tt #1}}}}

 \renewcommand{\mylabel}[1]{\label{#1}}
 
\newcommand{\mypict}[2]{\includegraphics[width=\figwidth#1]{images/#2}}

\newcommand{\Revision}{$Revision: 1.1 $}
\newcommand{\Thisrevision}{\Revision, Macros: \MacroRev}
\begin{document}

\preprint{\Thisrevision}

\title{Excitation Functions of the Analyzing Power in Elastic Proton-Proton  Scattering 
from 0.45 to 2.5 GeV} 
\authors

\institutes

\collaboration{EDDA Collaboration}

\date{\today \Thisrevision}

\Beginabstract{
Excitation functions $A_N(p_{lab},\Theta_{c.m.})$ of the analyzing power in elastic 
proton-proton scattering have been measured 
in an internal target experiment at the Cooler Synchrotron COSY
with an unpolarized proton beam and a 
polarized atomic hydrogen target. 
Data were taken continuously during the acceleration and deceleration for 
proton kinetic  energies $T_{lab}$ (momenta $p_{lab}$) 
between 0.45 and 2.5~GeV (1.0 and 3.3~GeV/c) and
scattering angles  30$^{\circ} \leq \Theta_{c.m.} \leq$ 
90$^{\circ}$. The results provide  excitation functions and angular 
distributions of high precision and internal consistency.
The data can be used as calibration standard between 0.45 and 2.5~GeV.
They have significant impact on phase shift solutions, 
in particular on the spin triplet phase 
shifts between 1.0 and 1.8 GeV.
}\Endabstract

\mypacs

\maketitle


\section{Introduction}

The internal target experiment EDDA 
\cite{albers97,Altmeier:2000,Albers:2004} 
at the Cooler Synchrotron COSY \cite{Maier:1997} is designed to provide
high precision measurements of proton-proton elastic scattering excitation
functions ranging from 0.45 to 2.5~GeV of laboratory kinetic energy $T_{lab}$.
In phase 1 of the experiment, spin-averaged differential cross sections were
measured continuously during acceleration with an internal 
polypropylene (CH$_2$) fiber target,
taking particular care to monitor luminosity as a function 
of beam momentum \cite{albers97,Albers:2004}.
In phase 2, excitation functions of the analyzing power $A_N$ 
are measured using a polarized atomic hydrogen beam as target.
In phase 3, excitation functions of the
polarization correlation parameters $A_{NN}$, $A_{SS}$ and $A_{SL}$
are measured using a polarized beam and a polarized target. 
The present paper gives a detailed account of our measurements 
of the analyzing power excitation functions  $A_N(p_{lab},\Theta_{c.m.}$)
using an internal polarized atomic beam target
during beam acceleration and decelaration, see also \cite{Altmeier:2000}.

Following the Argonne notation 
we denote the analyzing power by $A_N$  
\cite{Bourrely:1980mr} 
where the spin directions are defined
with respect to the laboratory frame of reference $(S,N,L)$ with 
$N$ the normal  to the scattering plane,
$L$ the longitudinal (beam) direction  and
$S$ the sideways direction ($\vec{S}=\vec{N}\times \vec{L}$).
In the  Saclay notation \cite{bystricky78} the laboratory frame of
reference is denoted $(s,n,k)$ with $(s,n,k){=}(S,N,L)$.
The analyzing power measured with beam polarization  
normal to the scattering plane is denoted
$(N,0;0,0){=}A_{00n0}$. If the target is polarized
the analyzing power is denoted
$(0,N;0,0){=}A_{000n}$.
In $pp$ elastic scattering beam and target particles are
identical and 
$A_{00n0}{=}A_{000n}$ =$A_N$. 
Often, the symbol $A$ or $P$ is  simply used for $A_N$.
Since
$A_N(\Theta_{c.m.})$ is antisymmetric with respect to
$\Theta_{c.m.}{=}90^{\circ}$,
the $A_N$ data are presented  only in the forward c.m. angle
range $0^{\circ}$--$90^{\circ}$. 


Elastic $pp$ scattering experiments 
\cite{Lechanoine:1993}
are fundamental
to the understanding of the NN interaction.
For kinetic energies below 1~GeV
a precise database of differential cross sections
and polarization observables has been accumulated.
These data are well represented by 
phase shift solutions 
\cite{Stoks:1993tb,Bystricky:1987,Bystricky:1990,Nagata:1996tp,
arndt94d,Arndt:1997if,Arndt:2000xc}. 
Modern phenomenological and meson-theoretical
potential models 
\cite{Lacombe:1980dr,Stoks:1994wp,Wiringa:1995wb,Machleidt:1987hj,Machleidt:1996km} 
provide excellent descriptions of the
data up to the pion threshold.
Extending the meson-theoretical models to higher energies requires the
inclusion of inelastic channel contributions.
Using relativistic transition potentials and
restricting to the NN, N$\Delta$ and $\Delta \Delta$ channels
yields reasonable descriptions of the data up to about 1~GeV 
\cite{Machleidt:1987hj}.
Those models can be improved by including other nucleon resonances
than the $\Delta$. But at certain higher energies
the meson exchange model has to break down when the
hadron substructure reveals itself in a crucial way \cite{machleidt89,Eyser:2003}.
Besides meson exchange models
recent theoretical work is based on chiral perturbation theory
($\chi$PT) 
\cite{Weinberg:1991um,Ordonez:1996rz,Kaplan:1998we,Epelbaum:1998ka,
Epelbaum:1999dj,Kaiser:1997mw,Kaiser:1998wa,Walzl:2000cx,Kaiser:2001dm},
Skyrmion or Soliton models 
\cite{Jackson:1985bn,VinhMau:1985sc,Walhout:1992ek,Pepin:1996mp} and
quark model descriptions 
\cite{Oka:1981ri,Oka:1981rj,Faessler:1983yd,Maltman:1984st,
Myhrer:1988af,Takeuchi:1989yz,Fernandez:1993hx,Entem:2000mq,
Shimizu:1999cr,Wu:1998wu,Faessler:1999gu} of the NN interaction.
A recent review of the theoretical progress can be found in
\cite{Machleidt:2001rw}.

Elastic $pp$ scattering
at GeV energies is ideally suited to study the short
range part of the NN interaction.
At 2.5~GeV kinetic energy a four momentum transfer of up to 
1.5~GeV/c is reached corresponding to spatial resolutions of about
0.13~fm. 
The precise knowledge of the analyzing power
as a function of energy   
provides a  focus on heavy meson exchanges, especially on the role
of the $\omega$-meson with respect to 
the spin-orbit force.
Apart from the true meson-exchange
genuinely new processes might occur at small distances
involving the
dynamics of the quark-gluon constituents.
Another issue related to the quark-gluon dynamics is the
question of existence or nonexistence of dibaryons.
Various QCD inspired models
predict dibaryonic resonances with c.m. resonance energies $E_R$ 
ranging between 2.1 and 2.9 GeV. 
Not any resonance has been observed so far.

A complete listing of previous analyzing power measurements 
in the kinetic energy range 0.45--2.5~GeV
can be found in the SAID database \cite{Arndt:2000xc}.
Recent measurements at discrete kinetic energies $T_{lab}{>}1$~GeV
are from the SATURNE II facility 
\cite{Perrot:1987pv,Arvieux:1997vg,Allgower:1998ma,
Allgower:1999ad,Allgower:1999ac,Ball:1999cn}. 
At lower kinetic energies $T_{lab}{<}1.0$~GeV
high precision measurements of the analyzing power
at discrete energies
have been performed 
at LAMPF \cite{bevington78,Mcnaughton:1982bt,McNaughton:1990sj}
and recently 
at IUCF \cite{Haeberli:1997,Rathmann:1998,vonPrzewoski:1998ye,Lorentz:2000rf}.

A first attempt to measure excitation functions of the 
analyzing power using an internal target during beam acceleration
has been performed at KEK \cite{shimizu90,kobayashi94}.
However, data were taken only at one fixed lab-angle, $\Theta_{lab}=68^{\circ}$,
from 0.5 to 2.0~GeV using a polarized beam and an unpolarized target.
In this experiment two narrow structures have been observed near
$T_{lab}{=}632$~MeV ($p_{lab}{=}1259$~MeV/c), i.e. in the neighbourhood of 
a depolarizing imperfection  resonance with $\gamma G{=}3$ of the KEK ring.
These structures were not confirmed in an external target experiment
at SATURNE II \cite{Beurtey:1992bw}.

The motivation of the EDDA experiment was to
measure excitation functions at intermediate
energies in a large angular range
with a high relative accuracy.
As a first result 
EDDA provided \cite{albers97} excitation 
functions $d\sigma/d\Omega(p_{lab}, \Theta_{c.m.})$ of  
unpolarized $pp$ scattering.  
Addition of these internally consistent data to the 
SAID data\-base \cite{arndt94d} allowed to 
extend the global PSA from 1.6 GeV to 2.5 GeV \cite{Arndt:1997if}.

The technique applied by EDDA 
to measure excitation functions during the acceleration of a proton beam
is perfectly appropriate 
to provide a new polarization standard in the form of precise 
excitation functions of the analyzing power $A_N$.
In addition, the measurement of analyzing powers
is a first necessary step in the EDDA program
towards measuring spin correlation parameters $A_{NN}$, $A_{SS}$ and
$A_{SL}$ as a function of energy.
With a polarized hydrogen target,
a high and stable
polarization is available and internally consistent analyzing power
data can be taken over a wide energy range.
In sect. II we describe the experimental setup and the modifications
of the EDDA detector to meet the increased demands for vertex reconstruction.
The data acquisition and processing is presented in sect. III.
The results are given in sect. IV and discussed in sect. V.


\section{Experimental Setup}

\subsection{Overview}

The EDDA experiment is designed to provide high precision measurements of
the proton-proton elastic scattering excitation functions over a
wide energy range. Using an internal target, data taking proceeds during
the synchrotron acceleration ramp of COSY, so that a complete excitation function
is measured during each acceleration cycle. Statistical accuracy is
obtained by averaging over many thousand cycles (multi-pass technique).
This technique requires a very stable and reproducible operation of COSY.
The internal recirculating COSY beam provides beam intensities high enough
for use of a polarized atomic beam target. 
Typical values are $3\cdot 10^{10}$ unpolarized protons
in the ring,  recirculation frequencies of 1.20 -- 1.57 MHz
and target densities of $2\cdot 10^{11}$ hydrogen atoms/cm$^2$
yielding luminosities of 7.2 -- $9.4\cdot 10^{27}$ cm$^{-2}$s$^{-1}$.

The analyzing power measurements were performed using an unpolarized proton beam
and a polarized hydrogen target. This method  
is in contrast to the usual method of using a polarized proton beam
and an unpolarized hydrogen target. It has the advantage to
avoid all uncertainties and  systematic errors due to
depolarization resonances in the acceleration ramp.
The direction of the target polarization was changed from
cycle to cycle between $\pm x$ and $\pm y$ thus allowing
a proper spin flip correction of false asymmetries \cite{ohlsen73}.
The absolute value of the target polarization is constant during 
the time period of an acceleration cycle (15~s).
Small drifts of the absolute polarization
over periods of up to two weeks are taken into account.
Therefore, excitation functions can be measured with  a high relative accuracy.

\subsection{Detector}

\begin{figure}[t]
\begin{center}
\includegraphics[width=\figwidth]{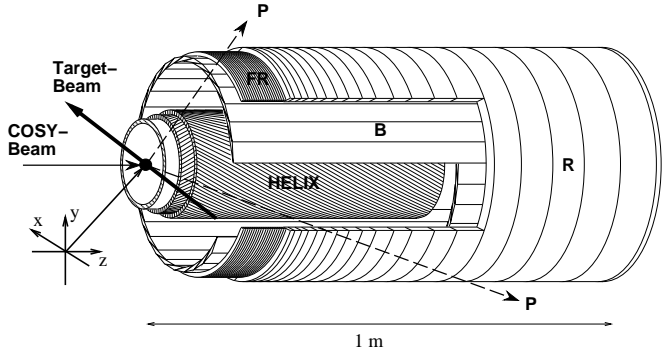}
\end{center}
\caption{Scheme of the EDDA detector. The outer hodoscope consists of
scintillator bars B, scintillator semi-rings R and semi-rings made of
scintillating fibers FR. The inner hodoscope HELIX consists of four layers of 
scintillating fibers helically wound in opposing directions.}
\label{fig:detector} 
\end{figure}

The EDDA detector shown in Figure \ref{fig:detector} in a schematic fashion consists 
of two cylindrical
detector shells.
The solid angle coverage is 30$^{\circ}$ to 150$^{\circ}$ in $\Theta_{c.m.}$ for
elastic proton-proton scattering and about 85 \% of 4$\pi $.
In phase~1 of the experiment only the outer detector shell was
used to take unpolarized differential cross section
data with a CH$_2$ target (and C target for background subtraction)
\cite{albers97,Albers:2004}.
The outer detector shell \cite{bisplinghoff93} 
consists of 32 scintillator bars (B) which are mounted parallel
to the beam axis. They are surrounded 
by scintillator semi-rings (R) and semi-rings made of scintillating
fibers (FR). The scintillator cross sections were 
designed so that each particle traversing the outer layers produces a
position dependent signal in two neighbouring bars and rings. The 
resulting polar and azimuthal angular resolutions are about 
1$^{\circ}$ and 1.9$^{\circ}$ FWHM.

For the measurement of spin observables, however, 
a polarized atomic beam target shall be used.
The interaction region of such a target is far from being pointlike.
The polarized hydrogen beam has a finite diameter of about 12 mm (FWHM).
It is superimposed on  
a background of unpolarized hydrogen atoms
which are due to the residual gas
in the beam pipe vacuum. Therefore,
a second inner detector shell is required to provide
an appropriate vertex reconstruction.

The inner detector shell (HELIX) is a cylindrical hodoscope
consisting of four layers of 2.5 mm diameter
plastic scintillating fibers which are helically
wound in opposing directions so that coincidence of hits
in the lefthanded and righthanded helices gives the point
at which the ejectile traversed the hodoscope.
The 640 scintillating fibers are connected to 16-channel multianode photomultipliers
and read out individually using the LeCroy 
proportional chamber operation system PCOS III.
A detailed description of the helical fiber detector can be found in
\cite{Altmeier:1999vx}. Its purpose is (i) vertex reconstruction of
elastic proton-proton scattering events in conjunction with the
outer shell, (ii) background suppression of scattering events from the
background of unpolarized hydrogen atoms surrounding the
polarized hydrogen beam and (iii) improved angular resolution.
Combined with the spatial 
resolution of the outer detector shell, the helix fiber detector provides for vertex 
reconstruction with a FWHM resolution of  1.3~mm 
in the x- and y-direction and 0.9~mm in the z-direction.
Using a fit of the vertex and scattering angles with
constraints imposed by $pp$ elastic scattering kinematics
the resulting polar and azimuthal angular resolutions are about 
0.3$^{\circ}$ and 1.3$^{\circ}$ FWHM.

\subsection{Polarized Atomic Hydrogen Beam Target}

\begin{figure}[bth]
\mypict{}{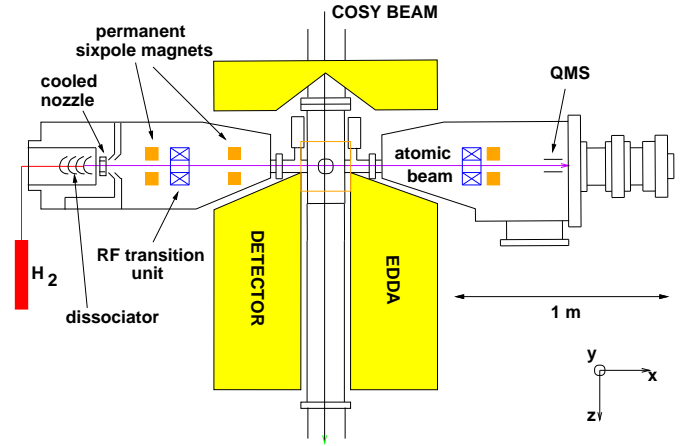}
\caption{Scheme of the polarized 
atomic hydrogen beam target.}
\label{fig2} 
\end{figure}

The polarized atomic hydrogen beam target 
\cite{Eversheim:1997ak}  is shown in Fig.~\ref{fig2}. 
Hydrogen 
atoms with nuclear polarization are prepared in an atomic-beam source
with dissociator, cooled nozzle, permanent sixpole magnets
and RF-transition unit.
The design of the atomic beam target had to meet constraints
imposed by the EDDA experiment. First, the space close to the
interaction region is limited: target components must be outside the
angular acceptance of the EDDA detector. This dictates a rather large 
distance of about 30~cm between the second sixpole magnet and the interaction region.
Second, 
in view of the closed orbit distortions in COSY
only weak guide fields are allowed such that only one pure hyperfine state can be used.
Third, the COSY beam width at injection  
and the change of the horizontal beam position during acceleration
makes the use
of a storage cell 
(typical apertures 10~mm x 10~mm) unfavorable. 
Reducing the beam width with beam scrapers,
and thus the injected beam current
would at least partly offset the benefit of higher target densities
and lead to increased background.

In the dissociator,
hydrogen is dissociated in an inductively coupled  350 W RF-discharge and passes through
an aluminum nozzle cooled to about 30~K and a skimmer. 
About $5\cdot 10^{16}$ hydrogen atoms per second are produced.
For the differential pumping a turbomolecular pump of 600 liter/s
is used in the first vacuum chamber between nozzle and skimmer and four
turbomolecular pumps of 1000 liter/s in the subsequent vacuum chambers.
In addition, a cryopump with a pumping speed of about 2300~liter/s is installed
in the second sixpole vacuum chamber. 
In the beam dump turbomolecular pumps of 360 and 600 liter/s and
a cryopump of 1000 liter/s are installed.
The comparable low temperature of
the nozzle leads to a decreased velocity (most
probable velocity 1.3~km/s) of the atomic beam and thus an increased target
thickness. The atomic beam target is usually operated at 0.5~mbar~liter/s
hydrogen flow. 
A small amount of oxygen flow is mixed
into the hydrogen flow yielding a 
thin layer of H$_2$O-molecules
in the region of the cooled nozzle.
Thus, the
recombination of
hydrogen atoms by the cooled nozzle is minimized
resulting in an increased target density.

The atomic beam source selects hydrogen atoms in a pure hyperfine state
($m_j$=$+1/2$, $m_I$=$+1/2$),
in order to achieve a high polarization in a weak magnetic holding field.
Here, $m_j$ and $m_I$ are the magnetic quantum numbers
of the electron and proton spins, respectively.
Hydrogen atoms in the $m_j$=$+1/2$ state are focused by the sixpole
magnets while those in the $m_j$=$-1/2$ state are defocused.
The first sixpole magnet removes the two hyperfine 
states with $m_j$=$-1/2$ and the Abragam-Winter RF-transition unit 
induces an intermediate field transition (IF-transition) 
to a depopulated hyperfine state, 
($m_j$=$+1/2$, $m_I$=$-1/2$)$\rightarrow$ ($m_j$=$-1/2$, $m_I$=$+1/2$), 
which is removed by the subsequent sixpole magnet.

\begin{figure}
\begin{center}
\includegraphics[width=\figwidth]{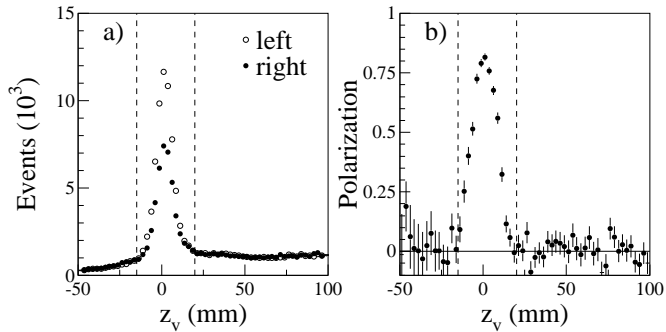}
\end{center}
\caption{(a) Elastic proton-proton scattering rates to the left and right of the
detector and (b) the derived polarization profile as a function of the
longitudinal vertex position $z_v$  measured with 
vertical target polarization at $T_{lab}{=}793$~MeV.
The polarized atomic hydrogen beam
stands out at $z_v{=}0$ on an unpolarized background
of the residual hydrogen gas in the beam pipe vacuum.
The dashed lines indicate the $z_v$  vertex cut [-15,+20]~mm.
The decreasing number
of events at $z_v{<}-15$~mm is due to the decreasing detector acceptance.}
\mylabel{fig:polprofile}
\end{figure}

In Fig. \ref{fig:polprofile} typical  
intensity and polarization distributions 
are shown as a function of the longitudinal position  $z_v$.
The plot shows the elastic proton-proton scattering
rates to the left and right of the detector (a)
and the derived polarization (b).
The polarized atomic hydrogen beam stands out at $z=0$
on an unpolarized background
which is due to the residual hydrogen gas in the beam pipe vacuum.
Polarizations of about 90~\% are deduced when the unpolarized background is subtracted.
The atomic beam width (FWHM) is about
12 mm   at the intersection 
with the COSY beam. 
A guide field in the interaction region of 1.0~mT
pointing up  or down 
was used to align the target spins
in the vertical direction.
The decrease in rate below $z{<}{-}15$~mm 
is due to the acceptance of the EDDA hardware trigger.
Including the unpolarized background, the effective polarization is about 73\%
for a $z$-vertex cut [-15~mm, +20~mm]  
and the effective target 
thickness is  1.8$\cdot$10$^{11}$ atoms/cm$^2$.  
The background  of unpolarized hydrogen depends on the quality of the
COSY vacuum. Therefore, in addition to the standard ion getter pumps,
titanium sublimation pumps are switched on.
Then, the resulting vacuum in the region of the EDDA-target
is  $<1\cdot 10^{-9}$~mbar without
and about $1\cdot 10^{-8}$~mbar with hydrogen beam target.

The atomic beam profile in $y$-direction has been measured at a fixed flattop energy 
(796~MeV) by sweeping the 
COSY beam by steerer magnets across the target and measuring the
vertex distribution of elastic proton-proton scattering events.
The target density distribution was deduced by fitting
a Gaussian distribution for the polarized atomic beam plus
a constant unpolarized hydrogen background.
The finite width of the COSY beam and
the detector resolution function were taken into account by 
an appropriate folding of the distributions.
The resulting atomic beam width (FWHM) in $y$-direction
of about 12~mm compares well with the corresponding width in $z$-direc\-tion. 
Similarly, the polarization profile in $y$-direction
which  was deduced from the 
measured left-right asymmetries compares well with the corresponding
profile in $z$-direction.

Due to the background of unpolarized hydrogen atoms
the effective polarization depends on the overlap of
the COSY beam with the polarized atomic beam.
Fortunately, variations of the vertical position and width of the COSY beam
during the acceleration are very small,
see Figs.~\ref{fig:xyprofile} and \ref{fig:beam_position_width}.
In addition, the COSY beam is much smaller
than the polarized atomic beam.
Therefore, the effective target polarization is
practically constant during the acceleration and deceleration.

In the beam dump the polarization of the
atomic beam is continuously monitored
using an additional 
RF-transition unit and a permanent sixpole magnet as Breit-Rabi polarimeter.

\subsection{Magnetic Guide Field}

\begin{figure}[t]
\begin{center}
\includegraphics[width=6.6cm,height=5cm]{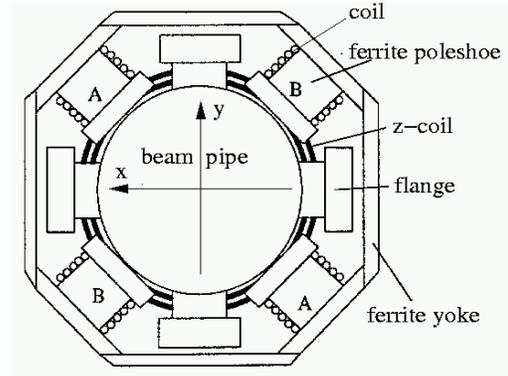}
\end{center}
\caption{Magnet for the guide field of the target polarization.
Magnetic guide 
fields of 1.0~mT in $\pm x$- and $\pm y$-direction are
produced near the beam axis 
by linear superposition of A- and B-fields.}
\mylabel{fig:magnet}
\end{figure}
The magnet for the guide field of the target polarization is shown schematically in
Fig.~\ref{fig:magnet}. It consists of a ferrite yoke and 
ferrite pole shoes located external to
the vacuum chamber.
The magnetic fields in $x$- and $y$-direction are produced by
superposing two fields of equal strength along the azimuthal directions 45$^{\circ}$
and 135$^\circ$. 
The earth magnetic field is shielded to a high degree
by the ferrite yoke. The residual magnetic field is
compensated by a small correction field using the
magnet for the guide field.
The currents
for the magnetic guide fields and the field corrections
are controlled by a personal computer.
The direction of the target polarization was changed from cycle to cycle 
by changing the direction of the guide field from $+x$ to $-x$ and $+y$ to $-y$. 
Using a flux gate probe the resulting field distributions were measured
as a function of $x$, $y$ and $z$.
These measurements proved a high magnetic field homogeneity 
across the fiducial vertex volume.

The strength of the magnetic field was chosen to be 1.0~mT. This value is sufficiently
large compared to the field distortion
produced by the circulating beam particles at the interaction point.
On the other hand, the resulting distortion of the closed orbit  is sufficiently small.
It can be calculated using the COSY lattice parameters at the target point.
For beam momenta between 1.0 and 3.3~GeV/c 
the resulting angle kick varies between 60 and 18~$\mu$rad
yielding a horizontal closed orbit shift $\Delta x$ between 51 and 16~$\mu$m for
guide fields along the $y$-direction.
Similar values are obtained for the vertical closed orbit distortions.
The closed orbit distortions deduced from the vertex reconstruction of the
data are in good agreement with these estimates.

\subsection{COSY Beam}

The ramping speed of COSY was changed
from the usual value 1.1~(GeV/c)/s to 0.55~(GeV)/s and
data were taken during acceleration as well as  
deceleration of the COSY beam. The time period of one cycle was about 15~s.
With an average of 2.8$\cdot$10$^{10}$ 
protons in the ring, luminosities of about  8$\cdot$10$^{27}$cm$^{-2}$s$^{-1}$  
were achieved and accumulated to an integrated luminosity of $\sim$10$^{34}$ cm$^{-2}$. 
The beam parameters were continuously measured during the
acceleration ramp. The beam momentum is derived from the RF of the cavity
and the circumference of the closed orbit with
an uncertainty of 0.25 to 2.0 MeV/c for the lowest and
highest momentum, respectively. 
\begin{figure}[h!]
\begin{center}
\includegraphics[width=\figwidth]{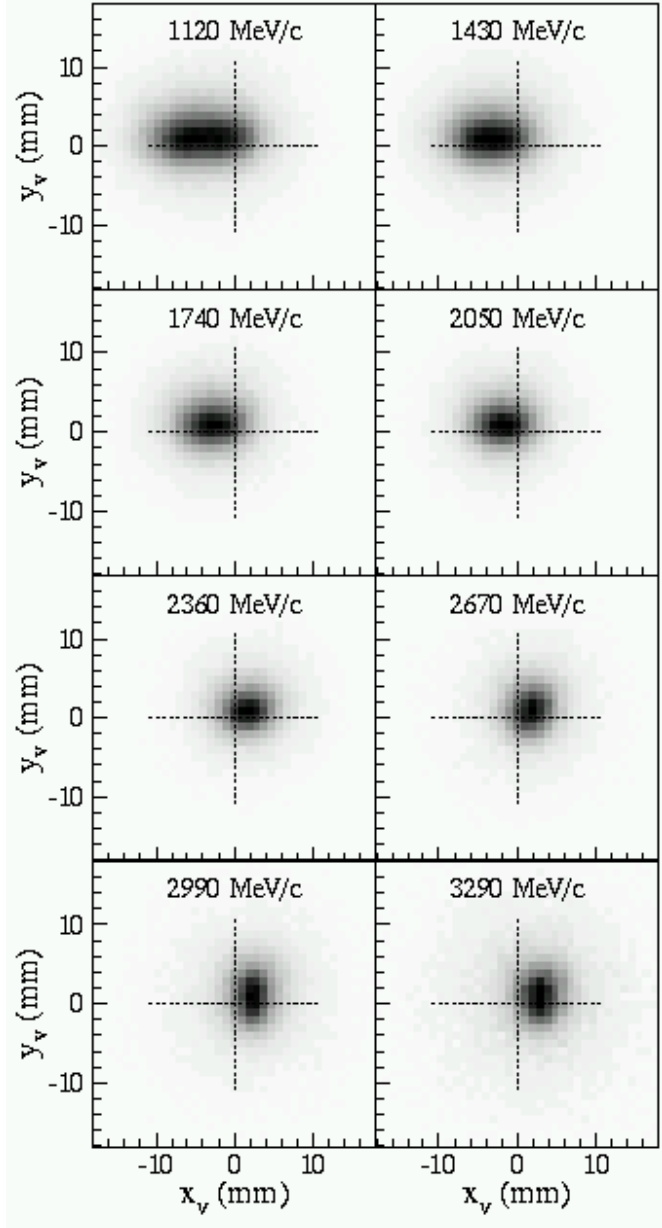}
\end{center}
\caption{Vertex distributions
in the $(x,y)$-plane for the running period November 98.} 
\mylabel{fig:xyprofile}
\end{figure}

The $x$-$y$ density distributions of the COSY beam 
can be deduced from the
measured vertex distributions. Thus, 
beam position and width can be reconstructed 
as a function of the beam momentum.  
Typical vertex
distributions are shown in Fig. \ref{fig:xyprofile} 
in the form of two-dimensional scatter plots.
Since the polarized atomic beam target is 
directed in $x$-direction and its width in $y$-direction is
rather large
these distributions resemble approximately the COSY beam density.
Interestingly, the horizontal beam position and width 
change slowly during the acceleration and deceleration
whereas the vertical beam position and width stay nearly constant.
This behaviour can be seen more clearly by
plotting the mean values and $\bar{x}$, $\bar{y}$  and the FWHM-widths 
$\Delta x{=}2.35 \sigma_x$, $\Delta y{=}2.35 \sigma_y$ 
as a function of the 
COSY cycle time $t$,
see Fig.~\ref{fig:beam_position_width}.

The deviations from the nominal beam position $\bar{x}=0$ and $\bar{y}=0$ 
indicate (i) misalignments of the detector and 
(ii) closed orbit distortions of the COSY beam
which are caused by small deviations from the nominal magnetic dipole fields in the 
accelerator ring.
In all three running periods the 
horizontal deviations 
$\bar{x}$
vary systematically as a
function of beam momentum between at most -5 mm and +2 mm.
Vertically, a small but constant offset of $\bar{y}\approx +1.5$~mm was 
observed, see Fig. \ref{fig:beam_position_width}.
\begin{figure}[t!]
\begin{center}
\includegraphics[width=\figwidth,bb= 104 302 500 730]{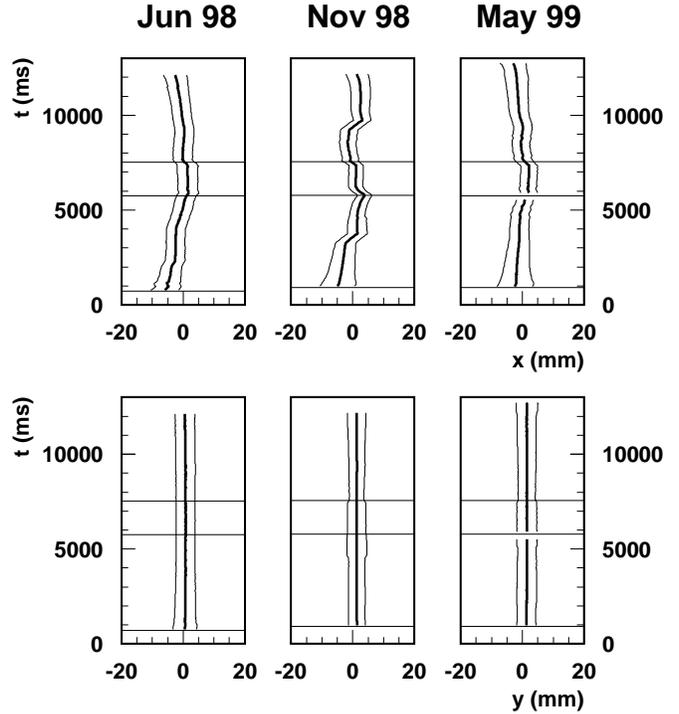}
\end{center}
\caption{Beam position (thick lines) and beam width (FWHM, thin lines)
in x- and y-direction
as a function of the accelerator time for three running periods.
The horizontal lines mark the start of data taking in the
acceleration ramp,
the beginning of the flattop momentum region and the beginning
of the deceleration ramp.}
\mylabel{fig:beam_position_width} 
\end{figure}

Variations of the horizontal and vertical beam widths $\sigma_x$ and $\sigma_y$
are expected since the adiabatic damping causes a $1/p_{lab}$ dependence of 
the beam emittances $\epsilon_x$ and $\epsilon_y$.  In addition 
the optics of the accelerator ring, 
i.e. the amplitude functions $\beta_x$ and $\beta_y$ may depend on $p_{lab}$.
The equations for $\sigma_x$ and $\sigma_y$ read
\begin{equation}
\sigma_x =\sqrt{\epsilon_{x} \beta_x},\; \; \; 
\sigma_y =\sqrt{\epsilon_{y} \beta_y}.
\end{equation}
If the amplitude functions $\beta_x$ and $\beta_y$ are constant
a $1/\sqrt{p_{lab}}$ dependence of 
the beam widths is expected.
However, in the chosen mode of operation   
$\beta_x$ was decreasing and $\beta_y$ increasing with $p_{lab}$.
As a consequence, the horizontal beam width $\sigma_x$ 
was rather strongly decreasing with $p_{lab}$
whereas the vertical beam width $\sigma_y$ was nearly constant.
\begin{figure}[t!]
\begin{center}
\includegraphics[width=\figwidth]{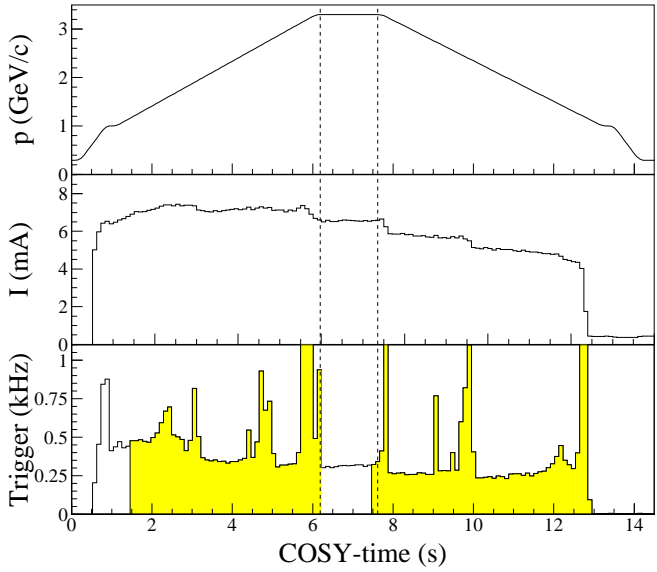}
\end{center}
\caption{
Typical COSY machine cycle: beam momentum $p{=}p_{lab}$,
beam current $I$ and trigger rate as a function
of the COSY cycle time. The dashed vertical lines
denote the flattop region.}
\mylabel{fig:cycle}
\end{figure}

The approximate constancy of the vertical beam profile
is of great importance for the effective target polarization
which depends on the overlap between  beam and  target.
The vertical width of the target beam (12~mm, FWHM)  
is about a factor two larger 
than the vertical width of the COSY beam
and the variations of the vertical beam position
and beam width 
during the acceleration and deceleration  are 
negligibly small with respect to the vertical width of the
target beam.

\section{Data Acquisition and Processing}

\subsection{Measuring Cycle}
Measurements of the excitation functions $A_{N}$($p_{lab}$, $\Theta_{c.m.}$) were 
performed in cycles of about 15~s duration with data 
acquisition extending over the acceleration, the flattop at 3.3~GeV/c, 
and the deceleration as well. 
A typical machine cycle is shown in Fig.~\ref{fig:cycle}.
The beam current is nearly constant during the acceleration
and slowly decreasing during the deceleration.
The trigger rate and the dead time show huge excursions at certain COSY-times
which are caused by increased background due to beam losses.
Fortunately, these excursions occured at different COSY-times
in the three running periods.
The direction of the target polarization was changed cyclewise yielding the
sequence $+x$, $-x$, 
$+y$, $-y$. The target was operating very stable and with constant 
polarization during subsequent acceleration cycles.

\subsection{Identification of Elastic pp Events}

The outer detector shell provides a fast and efficient trigger
based on (i) the coplanarity and (ii) the kinematic correlation
of two-prong events that fulfill the kinematics of elastic proton-proton scattering. 
Elastic proton-proton
scattering events are identified by coplanarity with the beam axis
\begin{equation}
|\varphi_1 - \varphi_2|=180^{\circ}
\end{equation}
and kinematic correlation
\begin{equation}
\tan \Theta_1 \tan \Theta_2 =\frac{2 m_p}{2 m_p +T_{lab}}.
\end{equation}
Here, $\Theta $ and $\varphi $ are the polar and azimuthal angles in the lab system,
$m_p$ is the mass of the proton and $T_{lab}$ its laboratory kinetic energy.
\begin{figure}[t!]
\begin{center}
\includegraphics[width=\figwidth]{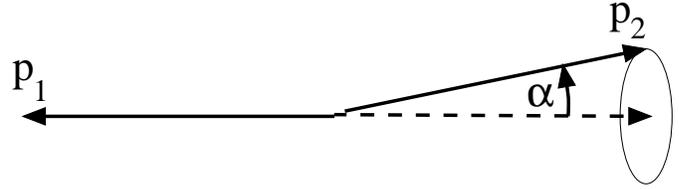}
\end{center}
\caption{Definition of the kinematic deficit $\alpha$ in the c.m. system.}
\mylabel{fig:kindef}
\end{figure}
In contrast to measurements of d$\sigma$/d$\Omega$ with CH$_2$ fiber targets 
\cite{albers97,Albers:2004}
the data rate with an atomic beam target was rather low.
Therefore, only the coplanar trigger, i.e. the coincidence of two
opposite bars, was applied in the on-line trigger
for the analyzing power measurements.
The kinematic correlation  was established in the off-line analysis.

The signature of an elastic event can be represented by 
one variable, the so-called
kinematic deficit $\alpha$, which gives the spatial angle deviation
from back-to-back scattering in the c.m.-system (see Fig.~\ref{fig:kindef}).
The kinematic deficit $\alpha$ can be determined
in the off-line analysis by transforming the
trajectories of two coincident particles into the c.m.-system
assuming the kinematics of elastic proton-proton scattering.
The resulting distributions of the spatial angle $\alpha$
start with zero at $\alpha=0^{\circ}$ and show a narrow peak
followed by a long tail, see Fig.~\ref{fig:simulation}. The finite width of the
elastic peak is due to the effects of small angle scattering and the finite
angle resolution of the detector. 
Elastic $pp$ scattering events can be identified using a momentum dependent cut,
\begin{equation}
\alpha < \alpha_{cut},\; \; \; \alpha_{cut}=[11.0-p/(1\;{\rm GeV/c})]^{\circ}.
\mylabel{alphacut}
\end{equation} 
\begin{figure}[t]
\begin{center}
\includegraphics[width=\figwidth]{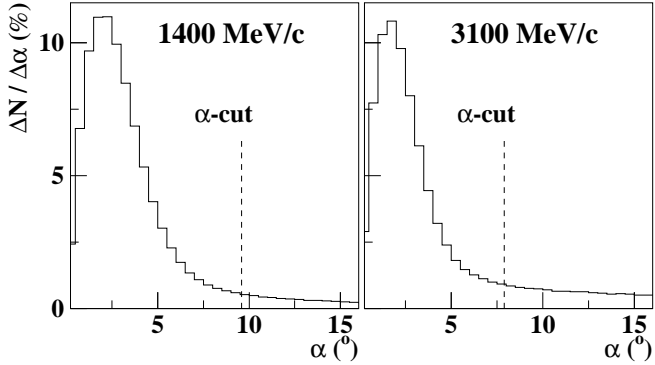}
\end{center}
\caption{Measured distributions of the kinematic deficit
$\alpha$ at two beam momenta. The vertical lines show 
the momentum dependent cut for the selection of elastic pp scattering
events.}
\mylabel{fig:simulation}
\end{figure}

\subsection{Vertex Reconstruction}

In the off-line analysis the coordinates of two kinematically correlated
trajectories are deduced from the hit pattern in the inner and outer detector shells.
The vertex is determined geometrically by the
subroutine FINDTRACKS  as the point of closest approach of 
two outgoing trajectories in the target region. 

In order to improve the vertex reconstruction
the  data are reanalyzed in a kinematic vertex fit 
with the kinematic constraints
of elastic proton-proton scattering.  
The kinematic fit of two proton trajectories to the hit pattern 
of an event under the constraint of 
elastic scattering kinematics is used to define a $\chi^2$ criterion for a further 
event selection. 
This method yields for instance 
at $T_{lab}{=}1500$~MeV and $\Theta_{c.m.}{=}40^{\circ}$  a vertex resolution with
$\sigma_x {=} 0.6$~mm, $\sigma_y {=} 0.6$~mm and $\sigma_z {=} 0.4$~mm
and an angle resolution with 
$\sigma_{\Theta_{c.m.}} {=} 0.6^{\circ}$ and $\sigma_{\varphi} {=} 0.8^{\circ}$.

The $\chi^2$-distribution is equivalent to the
$\alpha$-distribution with respect to the identification of elastic 
scattering 
events. 
The upper part of Fig.~\ref{fig:chi2} shows an
example of a $\chi^2$-distribution from a kinematic vertex fit
at $p_{lab}{=}2280$~MeV/c and $\Theta_{c.m.}{=}89^{\circ}$.
The dashed line is an extrapolation in order to estimate
an upper limit of
the background contribution. 
A momentum dependent cut
\begin{equation}
\chi^2<\chi^2_{cut},\; \; \; \chi^2_{cut}=28.0-5.5 p/{\rm GeV/c} 
\end{equation}
was chosen as final selection criterion for
an elastic $pp$ scattering event, see vertical lines 
in Fig.~\ref{fig:chi2}.

\subsection{Background}

\begin{figure}[t]
\begin{center}
\includegraphics[width=7cm]{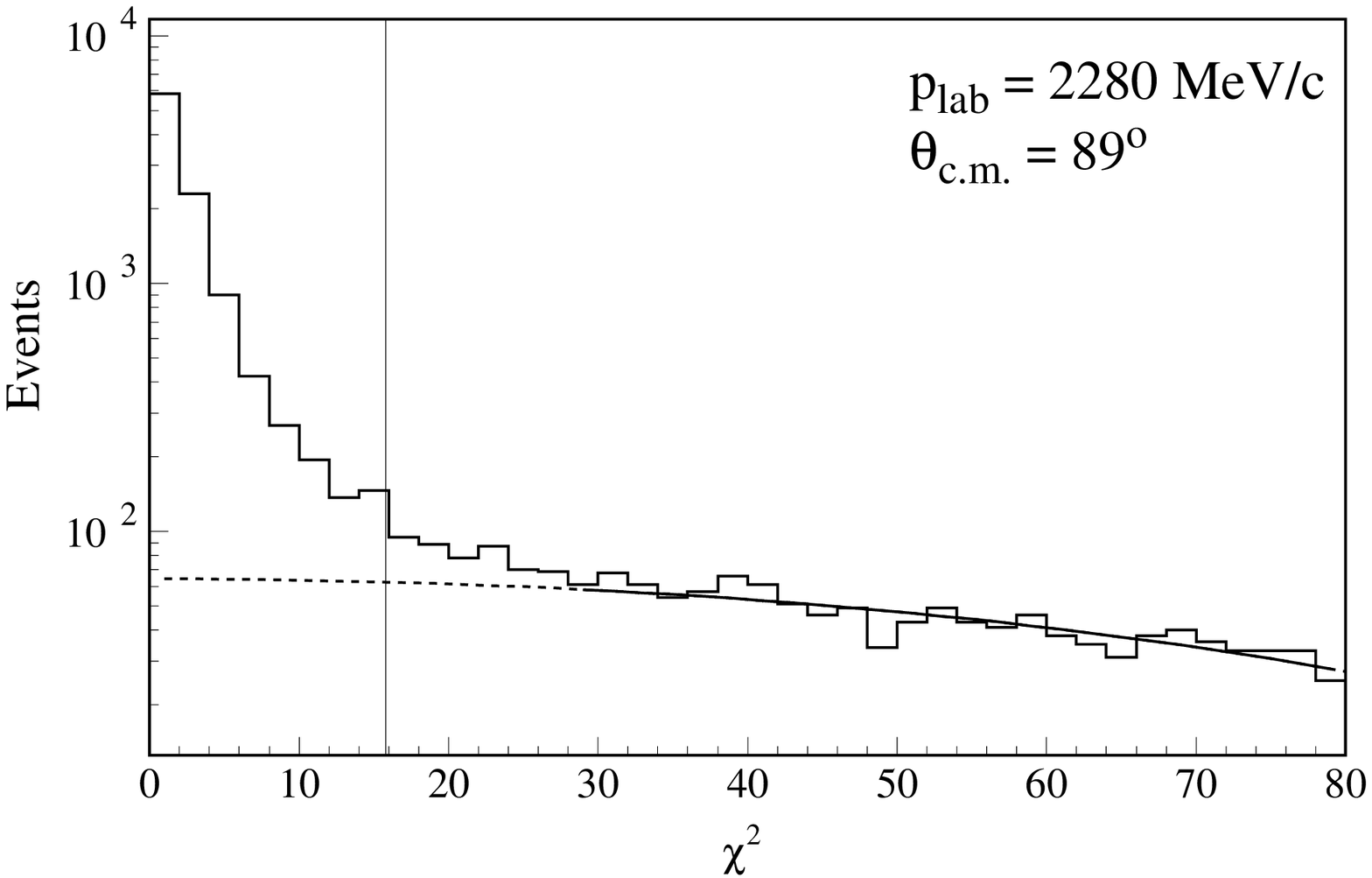}
\includegraphics[width=\figwidth]{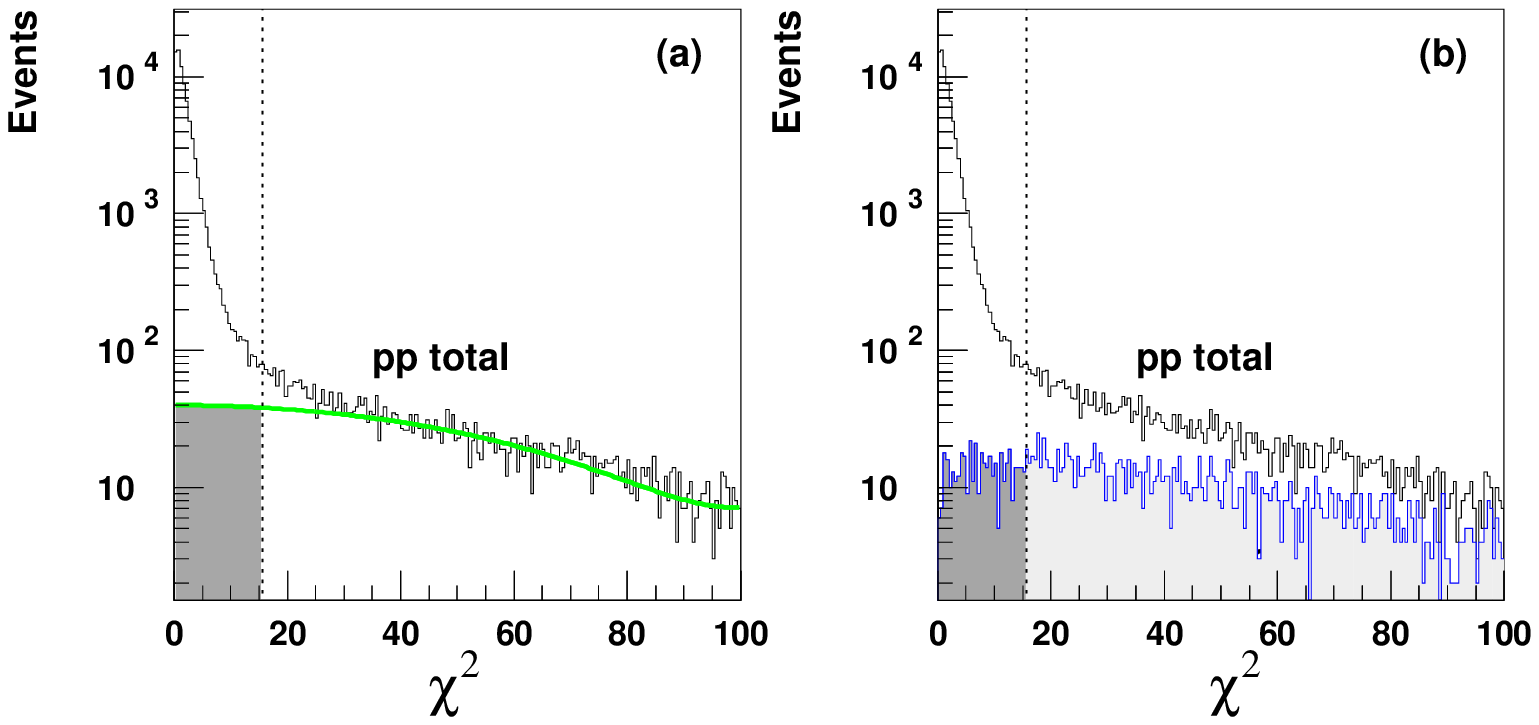}
\end{center}
\caption{Upper part: Example of a $\chi^2$-distribution.
The dashed line is an extrapolation from $\chi^2{>}30$ to zero. The vertical line shows 
the cut for the selection of elastic pp scattering
events. 
Lower part:
Monte Carlo simulation of a $\chi^2$-distribution at 2.3~GeV/c.
(a) Extrapolation from $\chi^2{>}30$ to zero.
(b) The inelastic background contribution (lower histogram)}
\mylabel{fig:chi2}
\end{figure}
The reduction of background takes advantage of the reconstructed vertex and the 
multiplicity patterns in both detector layers. Narrow cuts were applied to the 
hit pattern and to the vertex coordinate $z$ in COSY beam  direction, 
[-15 mm, +20 mm], a wider one 
in the $x,y$ plane around the beam profile (3$\sigma$ limits) along an ellipse 
following the slow drift and shape variation of the COSY beam during
acceleration,
see Figs.~\ref{fig:xyprofile},\ref{fig:beam_position_width}. 
The remaining inelastic background was estimated
 guided by Monte Carlo 
simulations of elastic and inelastic $pp$ interactions.
For the simulation of the hadronic
reactions 
the code MICRES \cite{Ackerstaff:2002hj} was used.

Monte Carlo simulations show that the long tail of the
$\alpha$- and $\chi^2$-distributions is mainly due to
misidentified elastic scattering events
suffering from a secondary reaction
in the beam pipe and the inner detector shell. 
This is in accordance with the fact
that events with $\alpha {>} \alpha_{cut}$ ($\chi^2 {>} \chi_{cut}^2$)
show  analyzing powers very similar
to the elastic scattering.
Therefore, extrapolating the $\alpha$- and $\chi^2$-distributions
from large $\chi^2$-values to zero
as shown in Fig.~\ref{fig:chi2}
overestimates  the inelastic background
under the elastic peak.
This background of inelastic reactions like
$pp\rightarrow pp\pi^0$, $pp\rightarrow pn\pi^+$, 
$pp\rightarrow pp\pi^+\pi^-$, $pp\rightarrow pp\pi^0\pi^0$,
and $pp\rightarrow pn\pi^+\pi^0$
is rather small.
It was estimated to be mostly $\leq$ 2\%  
and only at highest energies near $\Theta_{c.m.}{=}90^{\circ}$  up to 4.5\%.

\subsection{Determination of Analyzing Power}

\begin{figure}[t]
\begin{center}
\includegraphics[width=6cm]{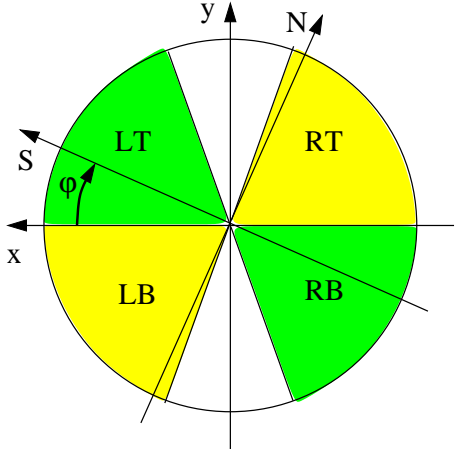}
\end{center}
\caption{
Definition of four detector sectors
for the determination of the analyzing power
with, looking along the beam axis, 
L~(Left), R~(Right), B~(Bottom) and T~(Top).
The azimuthal angle $\varphi$ gives the rotation of the
scattering plane and it's coordinate system $(S,N,L)$
with respect to the fixed coordinate system $(x,y,z)$ of beam and target.
The sectors  with $75^{\circ}{<}\varphi {<} 105^{\circ}$ and
$255^{\circ}{<}\varphi {<} 285^{\circ}$ are not used 
since the angle reconstruction is affected by the
readout of the half rings, see Fig.~1.}
\mylabel{fig:sector}
\end{figure}
We denote the 
target polarization by $\vec{Q}{=}(Q_x,Q_y,Q_z)$.
In order to eliminate systematic errors 
the direction of the target polarization was changed
from cycle to cycle between $\pm x$ and $\pm y$,
\begin{eqnarray}
+x&:&\vec{Q}=(+|\vec{Q}|,0,0), \; \; Q^+_x=+|\vec{Q}| \nonumber \\
-x&:&\vec{Q}=(-|\vec{Q}|,0,0), \; \; Q^-_x=-|\vec{Q}| \nonumber \\
+y&:&\vec{Q}=(0,+|\vec{Q}|,0), \; \; Q^+_y=+|\vec{Q}| \nonumber \\
-y&:&\vec{Q}=(0,-|\vec{Q}|,0), \; \; Q^-_y=-|\vec{Q}|. 
\end{eqnarray}
The polarized differential cross section may be written
\begin{equation}
\frac{{\rm d}\sigma}{{\rm d}\Omega}=
\frac{{\rm d}\sigma_0}{{\rm d}\Omega}
[1+A_N(Q_y\cos{\varphi}-Q_x\sin{\varphi})].
\end{equation}
Here, ${\rm d}\sigma_0/{\rm d}\Omega$ is the unpolarized 
differential cross section, $A_N$ the analyzing power
with respect to a polarization component in the direction
N normal to the scattering plane. The azimuthal angle $\varphi$
gives the rotation of the scattering plane and it's coordinate system
$(S,N,L)$ 
with respect to the fixed coordinate system (x,y,z) of beam and target,
see Fig.~\ref{fig:sector}.

In order to eliminate false asymmetries arising from
differences of the luminosities, efficiencies and solid angles
of the detector and from misalignments,
the geometric mean method of Ohlsen and Keaton \cite{ohlsen73} is used.
Two sets of cycles with opposite polarizations, e.g. $Q^+_{y}$ and 
$Q^-_{y}$, were combined to apply a ``proper spin flip''
correction for false asymmetries.  
Indicating the number of events obtained simultaneously
for the detector elements in the angle position $(\Theta,\varphi)$ 
with $N^+(\Theta,\varphi)$ for spin up and 
$N^-(\Theta,\varphi)$ for spin down
we define
\begin{equation}
L=\sqrt{N^+(\Theta,\varphi)N^-(\Theta,\varphi+\pi)}
\end{equation}
and
\begin{equation}
R=\sqrt{N^-(\Theta,\varphi)N^+(\Theta,\varphi+\pi)}
\end{equation}
to calculate the left--right asymmetry
\begin{equation}
\epsilon_{LR} = \frac{L-R}{L+R}
\end{equation}
for a pair of detector elements in the azimuthal 
positions $\varphi$ and $\varphi + \pi$
with $-\pi/2 < \varphi<+\pi/2$.
The analyzing power $A_N(\Theta)$ is deduced
as weighted mean over all $\varphi$-bins 
using
\begin{equation}
A_N(\Theta) = \frac{1}{\cos\varphi}\frac{\epsilon_{LR}(\Theta,\varphi)}{Q_y},
\; \; \; -\pi/2 < \varphi < \pi/2.
\end{equation}
Here, the moduli of the target polarizations are assumed to be equal, 
$|Q^+_y|=|Q^-_y|=|\vec{Q}|$.

Similarly the runs with $Q^{\pm}_{x}$ were used to deduce $A_N$ from the bottom--top 
asymmetries $\epsilon_{BT}$
for a pair of detector elements in the azimuthal positions
$\varphi$ and $\varphi + \pi$ 
with $-\pi < \varphi < 0$,
\begin{equation}
A_N(\Theta) = -\frac{1}{\sin\varphi}\frac{\epsilon_{BT}(\Theta,\varphi)}{Q_x},
\; \; \; -\pi < \varphi < 0,
\end{equation}
assuming $|Q^+_x|=|Q^-_x|=|\vec{Q}|$.
The terms Left (L), Right (R), Bottom (B) and Top (T) always refer 
to the scattered proton detected at 
forward scattering angles with $\Theta_{c.m.} \leq 90^{\circ}$.

This method eliminates exactly all false asymmetries
that means asymmetries which would still be observed with no
target polarization.
Thus, the result is independent of relative detector efficiencies
and solid angles, since they do not vary with time over 
the period of two adjacent cycles.
It is also independent of 
time fluctuations in the beam current or target density as well as differences
in the integrated charge and target thickness.

Misalignments of the beam axis with respect to the detector
axis yield small deviations from the nominal scattering angles $\Theta_{c.m.}$
and small variations of the solid angles.
These misalignments depend on the  closed orbit distortions
during the beam acceleration and deceleration.
Again, there is exact cancellation of false asymmetries due to deviations of the
solid angles. But since the analyzing power $A_N$ depends on the scattering
angle $\Theta_{c.m.}$
the determination of the effective
scattering angle becomes an important experimental task.
Fortunately, the EDDA detector allows to reconstruct the 
scattering angle $\Theta$ for each event with
high accuracy. Thus, systematic errors from
these deviations are also avoided.

The geometric mean correction method 
assumes the moduli of the target  polarizations $Q^{\pm}_{ y}$
and $Q^{\pm}_{ x}$ to be equal. 
This assumption is very well fulfilled since 
the polarization vector follows adiabatically the direction of the
magnetic guide field and 
the spin flip is
realized by flipping the direction of the guide field.
Small deviations that may occur cause negligible effects.
For instance deviations with
$|Q^+_y|-|Q^-_y| \leq 0.02$ influence $A_N$  by at most $7\cdot 10^{-5}$.

\subsection{Absolute Target Polarization}
\begin{figure}[t]
\begin{center}
\includegraphics[width=7cm]{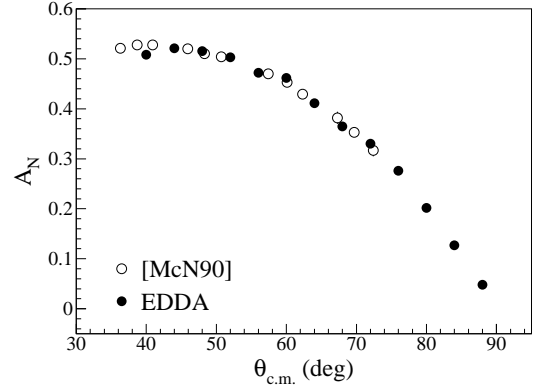}
\end{center}
\caption{
Angular distribution at 1379~MeV/c (730~MeV)
in comparison to the LAMPF data \cite{McNaughton:1990sj}
which were used for the absolute normalization.
The momentum bin width of the EDDA data is here 60~MeV/c.}
\mylabel{fig:normalization}
\end{figure}
The absolute values of the effective target polarizations 
$Q_x$ and $Q_y$ are established in each running period 
for {\bf one} momentum bin $\Delta p_{lab}$ = 60 MeV/c 
around the energy $T_{lab}=730$~MeV ($p_{lab} = 1379$~MeV/c). 
The absolute values of $Q_x$ and $Q_y$ are deduced as 
weighted means over all $\varphi$-bins from the measured asymmetries,
\begin{equation}
Q_x=-\frac{\epsilon_{BT}(\Theta, \varphi)}{\sin\varphi A_N(\Theta)},
\; \; \; -\pi < \varphi < 0,
\end{equation}
\begin{equation}
Q_y=\frac{\epsilon_{LR}(\Theta, \varphi)}{\cos\varphi A_N(\Theta)},
\; \; \; -\pi/2 < \varphi < \pi/2. 
\end{equation}
The precise angular 
distribution $A_N(\Theta)$ 
from McNaughton et al. \cite{McNaughton:1990sj} 
 with an absolute normalization error of 1~\% is taken as reference value,
see Fig.~\ref{fig:normalization}.
Taking the relative errors of the data into account the
overall normalization error of the present data is 1.2~\%.
The effective target polarization is constant during the 
acceleration and deceleration
since the overlap between COSY beam and polarized atomic beam target 
is constant.
The observed small variations of the vertical beam width 
(see Fig.~\ref{fig:beam_position_width}) cause
variations of the effective target polarization of less than 0.3~\%.

\subsection{Effective Target Polarization at Forward Angles}

The influence of the restrictions by the  detector acceptance 
has been studied separately. It yields small modifications
of the effective target polarization 
at low energies and small scattering angles. There,
events with negative $z$-values of the vertex point cannot be detected
if the recoil protons are outside the detector acceptance. 
As a consequence the effective target volume is reduced
in the $z$-direction from $[-15\; {\rm mm}, +20\; {\rm mm}]$
to $[z_{min},+20\; {\rm mm}]$ and the average polarization
within this reduced volume, subsequently called the effective
polarization, differs from the full-acceptance value. 
This modification 
is due to the background of unpolarized hydrogen atoms, see Fig.~\ref{fig:polprofile}.
It can be studied in a systematic way
by artificially restricting the vertex range in $z$-direction for
all $(\Theta_{c.m.},p_{lab})$-bins with full acceptance and deducing  the 
weighted mean of the resulting asymmetry ratios 
$\epsilon([z_{min},+20\; {\rm mm}])/\epsilon([-15\; {\rm mm}, +20\; {\rm mm}])$
as a function of $z_{min}$. 
These asymmetry ratios can directly be used as correction factors
of the effective target polarization at low energies and small scattering angles
where the $z$-range is restricted.

\section{Results}

\subsection{Consistency Checks}

The COSY beam changes its shape and position during acceleration and 
deceleration, though very reproducible in each cycle.
Prior to merging all 
data in one final set it was necessary to perform consistency checks on 
subsets obtained under different conditions of polarization and acceleration 
cycle. They demonstrated that 
the guide field is properly aligned to the detector coordinates (x,y), and that 
during acceleration and deceleration the same analyzing powers are obtained
\cite{Altmeier:2000}. 
This implies that vertex reconstruction and proper flip elimination of false 
asymmetries work well. 
A comparison of analyzing powers $A_N$ at $\Theta_{c.m.}{=}60^{\circ}$
acquired during the running periods June 98, November 98
and May 99 is shown in Fig.~\ref{fig:consistency}.
Since all data are compatible they are combined in 
one set. Altogether 3.1$\cdot$10$^7$ elastic scattering events were collected.
\begin{figure}[t]
\begin{center}
\includegraphics[width=\figwidth]{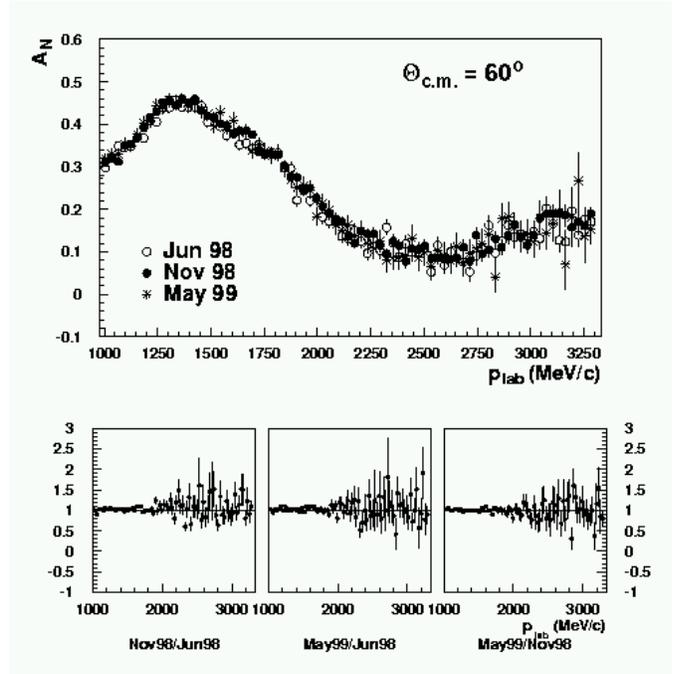}
\end{center}
\caption{Comparison of analyzing powers $A_N$ at $\Theta_{c.m.}{=}60^{\circ}$ 
acquired during the running periods June 98, November 98
and May 99.}
\mylabel{fig:consistency}
\end{figure}

\subsection{Errors}

Error estimates for $A_N$ include contributions from the maximum deviations between 
data subsets ($\leq$ 2.1\%), 
and the impact of the background on the asymmetry ($\leq$ 0.008). 
Errors 
due to closed orbit distortions by the
variation of the magnetic guide field with changes of the proton 
beam position and angle are negligible, see Sect.~2.4.
\begin{figure}[t!]
\begin{center}
\includegraphics[width=\figwidth]{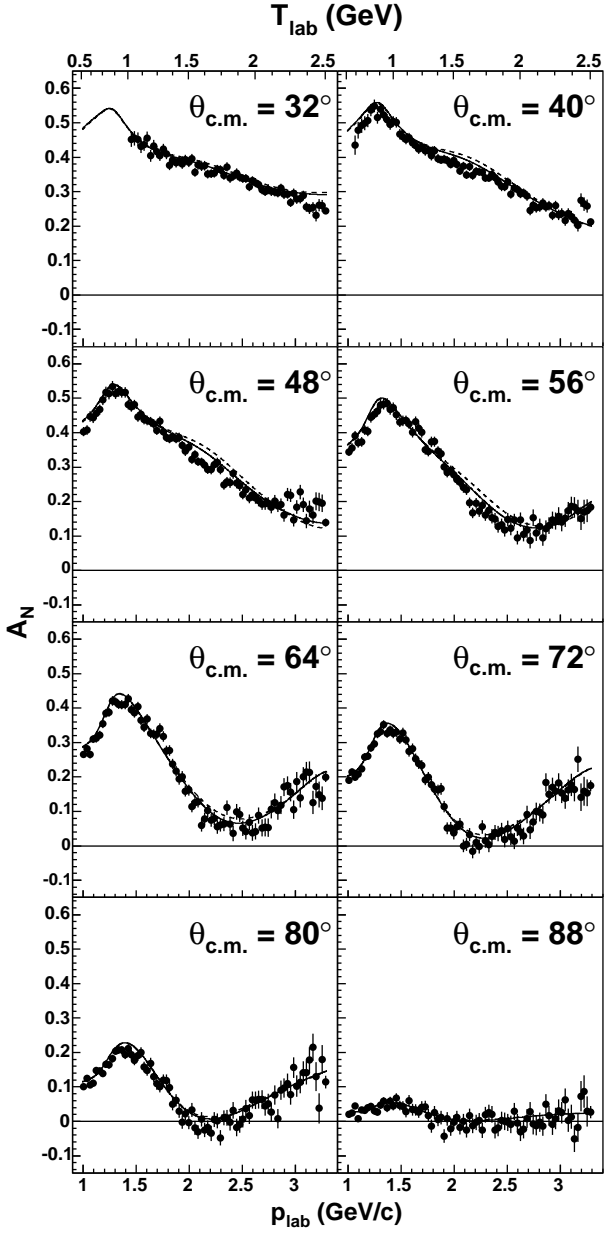}
\end{center}
\caption{
Excitation functions of $A_N$ from the present work at
eight out of 15 c.m. angles  
in comparison to the phase shift solutions SP00 (dashed curve) and  FA00 
(solid curve) \cite{Arndt:2000xc}.
The bin widths are  $\Delta p_{lab} {=}30$~MeV/c 
and $\Delta\Theta_{c.m.}{=} 4^{\circ}$.
The data at the flattop momentum 3.3~GeV/c exhibit an especially 
small statistical error.} 
\mylabel{fig:excitation_edda}
\end{figure}
The overall absolute normalization uncertainty 
of the excitation functions is $\pm 1.2$~\%
in the full
energy range 0.45--2.5~GeV. 
It is due to the absolute normalization uncertainty
of $\pm 1$~\% from the reference data \cite{McNaughton:1990sj} 
and the relative errors of the data at 730~MeV.
This systematic uncertainty is not
included in Figs.~\ref{fig:excitation_edda} and \ref{fig:angdist_edda} and the tables
available upon request \cite{DataAccess},
and must be applied to all data.

\subsection{Excitation Functions}

\begin{figure}[t!]
\begin{center}
\includegraphics[width=\figwidth]{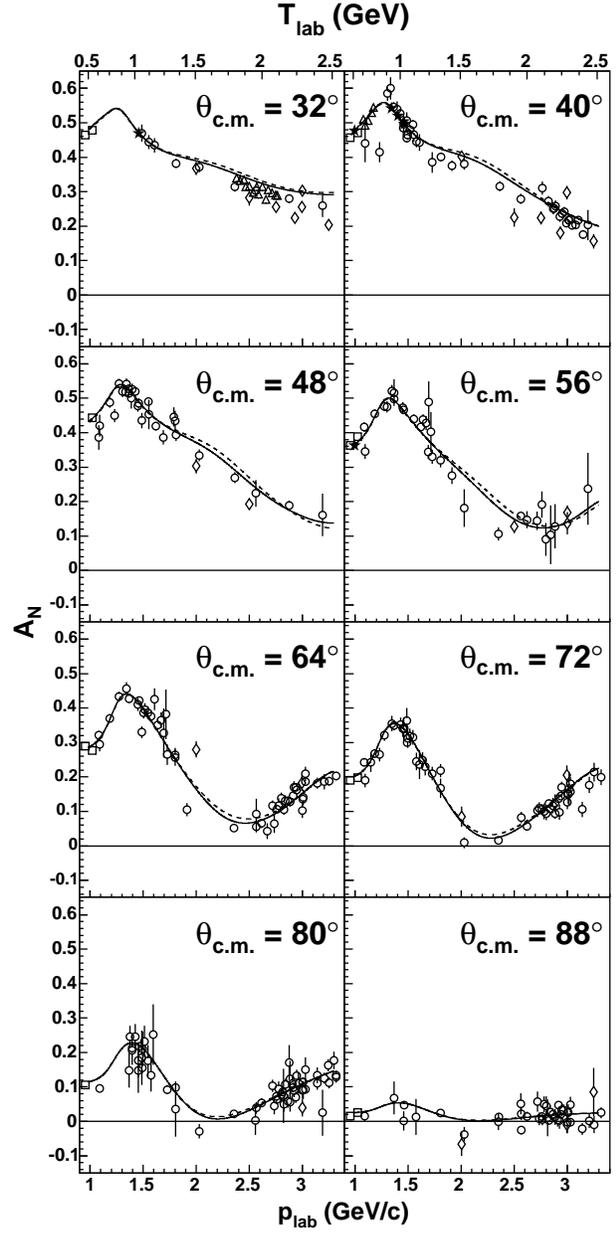}
\end{center}
\caption{Collection of published data from the data base of \cite{Arndt:2000xc}
without EDDA data of Altmeier et al. \cite{Altmeier:2000}
plotted as excitation functions at eight c.m. angles
in comparison to the phase shift solutions SP00 (dashed curve)  and FA00 (solid curve)
\cite{Arndt:2000xc}.
{\Large$\circ$} SATURNE II   \cite{Allgower:1999ad,Allgower:1999ac,
Ball:1999cn,Allgower:1998ma,Arvieux:1997vg,Perrot:1987pv}, 
$\triangle$ KEK   \cite{kobayashi94},
{\Large$\diamond$} ZGS   \cite{Parry:1973fj,Miller:1977pm,Bell:1980rp}, 
{\Large$\star$} LAMPF  \cite{McNaughton:1990sj,bevington78,Mcnaughton:1982bt},
$\square$ IUCF  \cite{vonPrzewoski:1998ye},
$+$ SIN \cite{Besset:1980sx,Berdoz:1983ih,Aprile:1986cx}.}
\mylabel{fig:excitation_other}
\end{figure}


Excitation functions $A_N(p_{lab}, \Theta_{c.m.}$) with  about 1150 data 
have been 
deduced from our experimental results by grouping them into 
$\Delta\Theta_{c.m.}{=} 4^{\circ}$ and $\Delta p_{lab}{=}30$~MeV/c 
wide bins. They 
supersede the results of the EDDA Collaboration reported
in   \cite{Altmeier:2000} and
are available upon request \cite{DataAccess}. 
Here, excitation functions at eight out of 15 c.m. angles 
are displayed in Fig.~\ref{fig:excitation_edda}.

\subsection{Angular Distributions}

\begin{figure}[t!]
\begin{center}
\includegraphics[width=\figwidth]{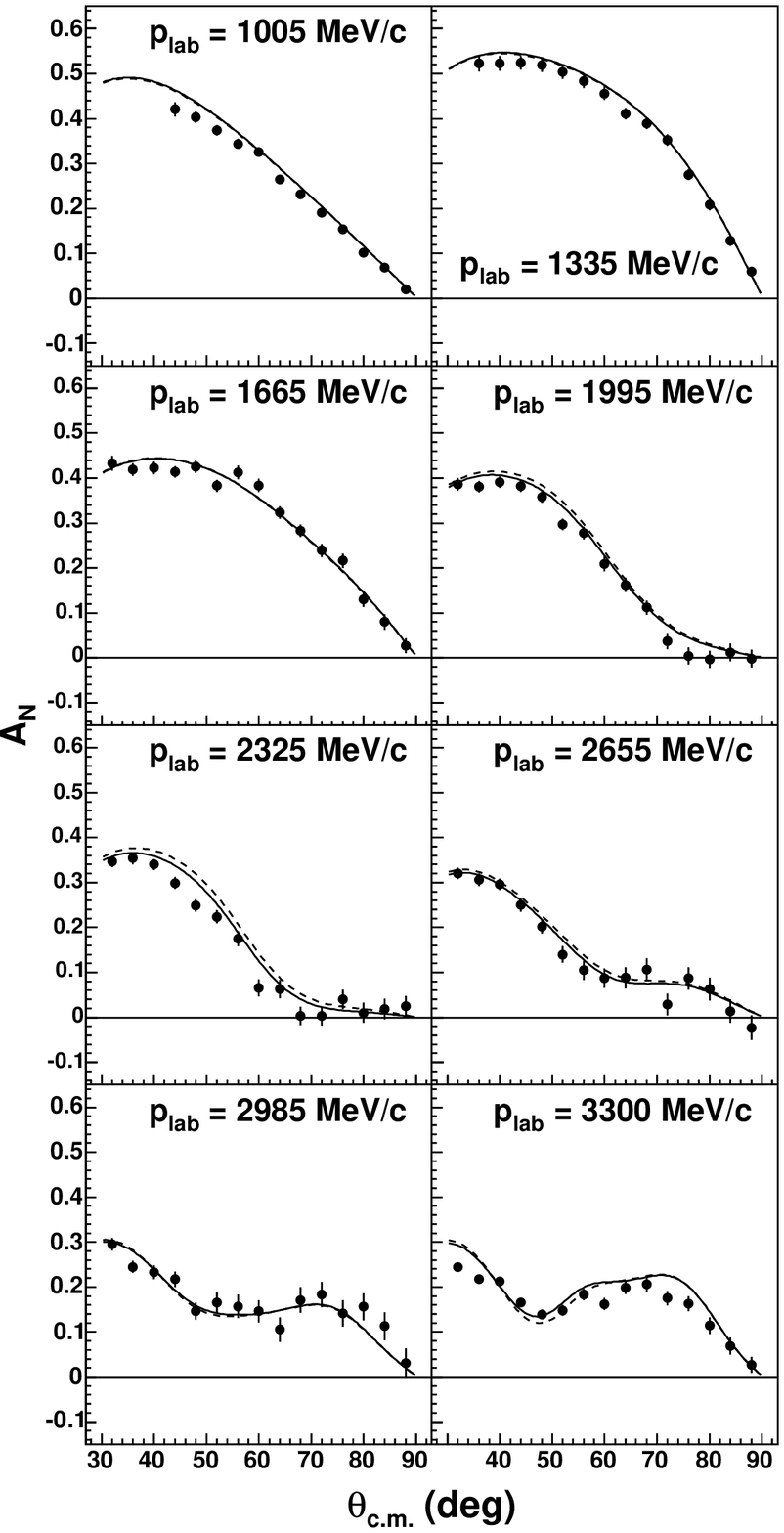}
\end{center}
\caption{
Angular distributions of $A_N$ from the present work at
eight out of 77 beam momenta  
in comparison to the phase shift solutions SP00 (dashed curve) 
and FA00 (solid curve) \cite{Arndt:2000xc}.
The bin widths are  $\Delta p_{lab}{=}30$~MeV/c 
and $\Delta\Theta_{c.m.} {=} 4^{\circ}$.
The data at the flattop momentum 3.3~GeV/c exhibit a 
small statistical error.} 
\mylabel{fig:angdist_edda}
\end{figure}
The data can also be presented in the form of angular distributions.
In Fig. \ref{fig:angdist_edda} eight out of 77 angular distributions
for $\Delta\Theta_{c.m.} {=} 4^{\circ}$ and $\Delta p_{lab} {=} 30$~MeV/c 
wide bins are shown. Again, these data   
supersede the results of the EDDA Collaboration reported
in   \cite{Altmeier:2000} and
are available upon request \cite{DataAccess}.

\begin{figure}[t!]
\begin{center}
\includegraphics[width=\figwidth]{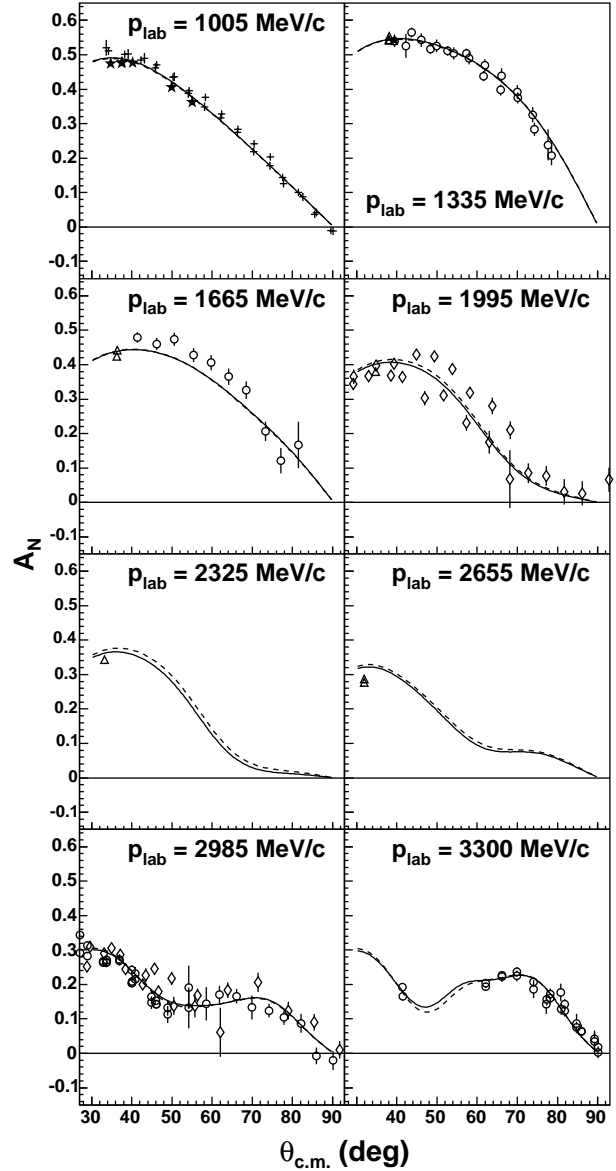}
\end{center}
\caption{Collection of published data from the data base of \cite{Arndt:2000xc}
without EDDA data of Altmeier et al. \cite{Altmeier:2000}
plotted as angular distributions at eight momenta
in comparison to the phase shift solutions SP00 (dashed curve)
and  FA00 (solid curve) \cite{Arndt:2000xc}.
{\Large$\circ$} SATURNE II   \cite{Allgower:1999ad,Allgower:1999ac,
Ball:1999cn,Allgower:1998ma,Arvieux:1997vg,Perrot:1987pv}, 
$\triangle$ KEK   \cite{kobayashi94},
{\Large$\diamond$} ZGS   \cite{Parry:1973fj,Miller:1977pm,Bell:1980rp}, 
{\Large$\star$} LAMPF  \cite{McNaughton:1990sj,bevington78,Mcnaughton:1982bt},
$\square$ IUCF  \cite{vonPrzewoski:1998ye},
$+$ SIN \cite{Besset:1980sx,Berdoz:1983ih,Aprile:1986cx}.}
\mylabel{fig:angdist_other}
\end{figure}

\section{Discussion}
\subsection{Comparison to other data}
Previous $A_N$ data were mainly measured at discrete energies.
A collection of published data from the SAID data base 
\cite{Arndt:2000xc}
is shown in Figs.~\ref{fig:excitation_other} 
and \ref{fig:angdist_other}.
The most recent results in the kinetic energy range 0.8--2.8~GeV
are from SATURNE II 
\cite{
Allgower:1999ad,Allgower:1999ac,Ball:1999cn,Allgower:1998ma,Arvieux:1997vg}.
They are in good agreement with the present
data. The same holds true for the results from SATURNE II
of Perrot et al. \cite{Perrot:1987pv} at the kinetic energies (momenta)
0.874 (1.550), 1.095 (1.804), 1.295 (2.027), 1.596 (2.354), 
1.796 (2.568), 2.096 (2.886) and 2.396~GeV (3.200~GeV/c).
The ZGS data of Parry et al. \cite{Parry:1973fj} at 
1.73 (2.5), 1.97 (2.75), 2.14 (2.93) and 2.44~GeV (3.25~GeV/c)
and Miller et al. \cite{Miller:1977pm} at 
1.27 (2.0) and 2.21~GeV (3.0~GeV/c) show considerable deviations
in the angular distributions.

In the kinetic energy range 0.45--0.8~GeV
the world data set exhibits
several high precision measurements at discrete
energies.
The absolute normalization of our $A_N$ excitation functions was established
at $T_{lab}{=}730$~MeV ($p_{lab}{=}1379$~MeV/c)
by taking the precise angular distribution $A_N(\Theta)$ of
McNaughton et al. \cite{McNaughton:1990sj} as reference value.
Our data below and above 730 MeV (1379~MeV/c) are in very good
agreement with other precise LAMPF measurements of
McNaughton et al. \cite{Mcnaughton:1982bt}
at 643~MeV (1273~MeV/c) and Bevington et al. \cite{bevington78}
at 787 (1448) and 796~MeV (1459~MeV/c)
as well as with precise IUCF measurements of Przewoski et al. \cite{vonPrzewoski:1998ye}
at 448.9~MeV (1022~MeV/c), SIN measurements of 
Besset et al.  \cite{Besset:1980sx}, 
Berdoz et al. \cite{Berdoz:1983ih}, 
Aprile et al. \cite{Aprile:1986cx} between 400  and 600~MeV (954 and 1219 MeV/c)
and SATURNE measurements  of Allgower et al. \cite{Allgower:1999ac}
at 795 MeV (1457~MeV/c).

\subsection{Comparison to Partial-Wave Analyses}

After the publication of the first unpolarized differential cross section
data from EDDA~\cite{albers97}  the VPI group extended their
energy dependent phase shift analysis from 1.6~GeV (2.36~GeV/c)
up to 2.5~GeV (3.3~GeV/c) laboratory kinetic energy (beam momentum)
with the solution SM97~\cite{Arndt:1997if}.
Meanwhile  energy dependent phase shift solutions
are available with a maximum beam energy (beam momentum)
of 3.0~GeV (3.82~GeV/c)~\cite{Arndt:2000xc}.
The solution SP00 includes the 
recent $A_N$ data from IUCF
\cite{Haeberli:1997,Rathmann:1998,vonPrzewoski:1998ye,Lorentz:2000rf} 
and Saclay \cite{Allgower:1999ad,Allgower:1999ac}.
The recent $A_N$ data from EDDA~\cite{Altmeier:2000} 
are included in the solution FA00.

The comparison of the present data to the phase shift solution 
FA00 \cite{Arndt:2000xc} yields 
agreement in the size and the general angle and momentum dependence  
of the excitation functions and angular distributions,
see Figs.~\ref{fig:excitation_edda} and \ref{fig:angdist_edda}.
Small systematic deviations can be seen in the excitation 
functions, in particular for momenta from 1.8 -- 2.5~GeV/c. 
As can be seen in Fig. \ref{fig:excitation_edda} the difference between 
the solutions FA00 and SP00 is small and the systematic deviations
in the momentum range 1800 -- 2500~MeV/c remain.
It is interesting to note that the analyzing power data 
are slightly negative in the region around $p_{lab}{=}2.0$~GeV/c and
$\Theta_{c.m.}{=}80^{\circ}$
whereas the the phase shift solutions remain positive. 
This observation is in agreement with
previous Saclay data \cite{Perrot:1987pv} in that region.
 
Including the EDDA data into the solution FA00 turned out that 
the agreement of the phase shift solution with the new 
$A_N$ data was slightly improved.
The spin triplet phases, 
e.g. those of the $^3F_2$ partial wave  experienced 
significant changes. 
This is due to the 
fact, that the analyzing power times differential cross section 
is equal to the real part $Re(a^{\star}\cdot e)$, where the invariant amplitudes 
$a$ and $e$ \cite{Lechanoine:1993,bystricky78} include only triplet partial waves.

\subsection{Sensitivity to narrow resonances}

All excitation functions of the analyzing powers 
show a smooth dependence on beam momentum.
There is no evidence for narrow resonances. 

A previous internal target experiment at KEK
\cite{kobayashi94} observed
two narrow structures in the excitation function of the
analyzing power $A_N$ at a laboratory kinetic energy
near 632~MeV corresponding to $\sqrt{s}=2.17$~GeV.
Those measurements were performed using a 
polarized proton beam in the KEK ring
and a 30~$\mu$m thick polyethylene fiber target.
The outgoing protons were detected at a fixed
backward angle of 68$^{\circ}$ in coincidence with
a forward detector. The corresponding c.m. angle near 632~MeV
was 38.4$^{\circ}$. 
In the present work all excitation functions
including those near 38$^{\circ}$ are smooth.
Our result is in agreement with the measurements of
Beurty et al. \cite{Beurtey:1992bw}.

The fact that $A_Nd\sigma/d\Omega=Re(a^{\star}\cdot e)$
includes only triplet partial waves 
implies that the excitation functions for 
$A_N$ are more sensitive to resonant excursions in triplet than to those in 
singlet partial waves.
However, the excitation functions measured here 
exhibit no evidence for energy--dependent structures.
It should be noted that also the excitation functions of
$d\sigma/d\Omega$ did not show any evidence for energy--dependent
structures \cite{albers97,Albers:2004}.
A more detailed discussion of sensitivities to and upper limits for such 
structures will be given in a forthcoming paper.

\section{Conclusions}

In conclusion, we report on the first measurement of  analyzing power
excitation functions 
$A_N(p_{lab}, \Theta_{c.m.})$ in the 
laboratory momentum range 1.0--3.3~GeV/c and the
c.m. angle range
$30^{\circ}$--$90^{\circ}$
for proton-proton 
scattering during acceleration and deceleration in a  
synchrotron. 
The data provide a new polarization standard and can be used for
calibration purposes in the full energy range 0.45--2.5~GeV.
The excitation functions agree with fixed energy data 
and close the  gaps in between with data of high precision 
and consistency. The phase shift analysis including our data yields a
global phase shift solution for $T_{lab}$ up to 2.5~GeV (FA00)  
showing distinct deviations from previous  
phase shift solutions that occur mainly in the spin triplet phases. 
Further progress can be expected from the new 
excitation functions of the spin correlation
parameters $A_{NN}$, $A_{SS}$ and $A_{SL}$
measured by the EDDA experiment. 

\section*{Acknowledgements}
The EDDA collaboration gratefully acknowledges the great support 
received from the COSY accelerator group. Helpful discussions with and 
comments from R. A. Arndt, Ch. Elster, and R. Machleidt are very much 
appreciated. This work was supported by the BMBF, Contracts 06BN910I and 
06HH152, and by the Forschungszentrum J\"ulich, Contracts FFE 41126903 and 41520732.

\printfigures
\bibliographystyle{unsrt}
\bibliography{rohdjess}

\end{document}